\documentclass[letterpaper, 10 pt, conference]{ieeeconf}  

\IEEEoverridecommandlockouts                              

\usepackage{cite}
\usepackage{amsmath,amssymb,amsfonts}
\usepackage{graphicx}
\usepackage{textcomp}
\usepackage{xcolor}
\usepackage{subcaption}
\usepackage{bm}
\usepackage[colorlinks=True, citecolor=red, linkcolor=blue]{hyperref}
\usepackage{booktabs}
\usepackage{orcidlink}
\usepackage{algorithm}
\usepackage{algpseudocode}
\usepackage{mathtools}
\usepackage{todonotes}
\usepackage{xcolor}

\newcommand{\ybi}[2]{\bar{\bm y}_{#1}^{#2}}
\newcommand{\yb}[1]{\bar{\bm y}^{#1}}
\DeclareMathOperator*{\argmax}{arg\,max}
\DeclareMathOperator*{\argmin}{arg\,min}

\def\BibTeX{{\rm B\kern-.05em{\sc i\kern-.025em b}\kern-.08em
    T\kern-.1667em\lower.7ex\hbox{E}\kern-.125emX}}

\title{\LARGE \bf
Distributed Nonlinear Filtering using Triangular Transport Maps\thanks{Approved for public release; distribution is unlimited. Public Affairs
approval \# AFRL-2023-5018}
}

\author{Daniel Grange$^{*}$\orcidlink{0000-0002-7643-496X} \and 
\and Ricardo Baptista$^{\dagger}$\orcidlink{0000-0002-0421-890X} \and Amirhossein Taghvaei$^{\ddagger}$\orcidlink{0000-0002-1536-892X}  \and 
Allen Tannenbaum$^{*}$\orcidlink{0000-0002-0567-5256} \and
Sean Phillips$^{\mathsection}$\orcidlink{/0000-0001-9074-6049}
\thanks{$^{*}$Department of Computer Science, Stony Brook University, NY; \texttt{\{daniel.grange,allen.tannenbaum\}@stonybrook.edu}.}
\thanks{$^{\dagger}$Department of Computing and Mathematical Sciences, California Insitute of Technology, Pasadena, CA; \texttt{rsb@caltech.edu}.}
\thanks{$^{\ddagger}$Department of Aeronautics and Astronautics, University of Washington, Seattle, WA; \texttt{amirtag@uw.edu}.}
\thanks{$^{\mathsection}$Space Vehicles Directorate, Air Force Research Laboratory, Kirtland AFB, NM; \texttt{sean.phillips.9@spaceforce.mil}}
}

\begin{document}

\maketitle
\thispagestyle{empty}
\pagestyle{empty}

\begin{abstract}

The distributed filtering problem sequentially estimates a global state variable using observations from a network of local sensors with different measurement models. In this work, we introduce a novel methodology for distributed nonlinear filtering by combining techniques from transportation of measures, dimensionality reduction, and consensus algorithms. We illustrate our methodology on a satellite pose estimation problem from a network of direct and indirect observers. The numerical results serve as a proof of concept, offering new venues for theoretical and applied research in the domain of distributed filtering.
\end{abstract}


\section{Introduction}

Multi-agent systems are commonplace in today's technological landscape, and many problems that were once cast in a centralized setting, have been recast in a distributed manner~\cite{boyd_distributed_2010}. With the introduction of multiple agents, various considerations must be made due to information flow, changes in network topology, and time-delays. Particularly, ensuring agreement among network agents is paramount. In aerospace, one may find a domain where multiple observations within a network of sensors aim to estimate a dynamic state. For example, satellites in orbit can freely position themselves along six degrees of freedom, and being able to recognize the pose in addition to the instantaneous change in pose is an essential priority in space-based missions such as docking and space debris cleanup~\cite{kisantal_satellite_2020, stacey_autonomous_2022}. 

To begin addressing these problems for nonlinear dynamical systems, filters have been developed to varying degrees of success. One such example, the ensemble Kalman filter (EnKF), is a widely used method that represents the distribution of state using a set of state vectors, called an ensemble~\cite{Tong2018}. Despite being inconsistent for non-Gaussian problems, the EnKF is robust in high dimensions. 
Alternatively, particle filters can offer a consistent numerical approximation to the nonlinear Bayesian filtering problem, and during the early 1990s it became the methodology of choice for certain classes of nonlinear and non-Gaussian systems \cite{hlinka_distributed_2013}. 
These algorithms seek to sequentially estimate the state of some system by recursively updating an ensemble of estimates, called ``particles''. 
In the case of distributed systems, this approach has been extended to so-called distributed particle filters~\cite{hlinka_distributed_2013, dehghanpour_survey_2019}. These approaches can be computationally inefficient, especially when the state space is high-dimensional, however, as particles degenerate and sampling becomes impoverished \cite{snyder_obstacles_2008}. Instead of updating weights on samples from a distribution, as is the case with sequential importance re-sampling (SIR) particle filters, 
measure transport offers a different approach to map samples between two transformations~\cite{taghvaei_optimal_2021, baptista_representation_2022, zhang_optimal_2023, chen_optimal_2021}. One such transformation is based on optimal transport, which minimizes an associated transport cost of moving the samples~\cite{santambrogio_optimal_2015}. 

One attractive instance of measure transport for Bayesian inference is through the approximation of the Knothe-Rosenblatt (KR) rearrangement~\cite{villani_optimal_2009, el_moselhy_bayesian_2012}. This transformation can be easily approximated given only samples of a distribution and has been applied for various higher-dimensional and nonlinear filtering problems~\cite{spantini_coupling_2022, le2022low}.
In addition, synthesizing multiple observations through the KR rearrangement has been explored by~\cite{kim_feedback_2021}, and in the context of robotics by~\cite{huang_nf-isam_2021}. 
The aim of this work is to construct a global transport map from which a network of agents is able to sequentially estimate a satellite's pose in a scheme that conforms to the limitations imposed by the space-domain.



To this end, we first formally state the problem of distributed nonlinear particle filtering in Section~\ref{sec:problem_formulation}. Section~\ref{sec:background} establishes the framework of measure transport and approximating the Knothe-Rosenblatt rearrangement. The novel design for adapting triangular transport maps filters to the multi-agent setting is introduced in Section~\ref{sec:data_fusion} and demonstrated with numerical results in Section~\ref{sec:case_study}.

\section{Problem Formulation} 
\label{sec:problem_formulation}
Suppose there is a sensory network comprised of $N$ agents that observe a global variable $\bm x(t_i) \in \mathbb{R}^n$ representing the state of a dynamical system at time $t_i$. Each agent $l,$ for $1 \leq l \leq N$, observes the state at irregularly spaced times $t_{i},$ for $i \in \mathbb{N}_0$, according to some noisy observation $\bm y^{l} \in \mathbb{R}^{d_l}$ of $\bm x(t_i)$. That is, the 
states and observations follow 
\begin{align}
    \text{State Model:}\quad & \bm x(t_{i}) = \bm f(\bm x(t_{i-1}), \xi_i)
    \label{eq:dynamics} \\
    \text{Observation Model:}\quad & \bm y^{l}(t_i) = \bm h^{l}(\bm x(t_i), \zeta_i^l),
    \label{eq:observation}
\end{align}
where $\xi_i$ and $\zeta_i^l$ are mutually independent process and observation noise variables,  respectively. 

The network topology of the agents is described by an adjacency matrix $A \in \mathbb{R}^{N \times N}$, where $A_{l,l'} = 1$ if agent $l$ can communicate information to agent $l'$, otherwise $A_{l,l'} = 0$. We assume the network connectivity to be constant over time. 
We also denote the neighbors of an agent as the set of indices $\text{Nbs}(l) = \{l' \in \mathbb{Z} | 1 \leq j \leq N, A_{l,l'} = 1 \}$. At each time, agent $l$ collects its neighbors observations as $\{\bm y^{l}\}_{l' \in \text{Nbs}(l)}$. We denote the concatenation of this sequence of observations as $\yb{l} \in \mathbb{R}^{\bar{d_l}}$ where $\bar{d_l} = \sum_{l' \in \text{Nbs}(l)}d_l$. 

In this work, we consider the problem of sequentially estimating the posterior distribution $\pi_{\bm x(t_i)|\bm y(t_0),...,\bm y(t_i)}$ for each agent where $\bm y(t_i)$ is the vector of length $d = \sum_{l=1}^N d_l$ that is constituted by a concatenation of measurements $\bm y^l(t_i)$ taken by all agents at time step $t_i$. More specifically, we wish to reconstruct the state using a distributed system of agents $1 \leq l \leq N$ that make partial observations through Bayesian inference. As with particle filters, we approximate the posterior with an ensemble of particles $\{\bm x_1^l,...,\bm x_M^l\}$ at each time step.

\section{Background}
\label{sec:background}






\subsection{Measure Transport}

Motivated by the recent successes of transport methods for applications in control and nonlinear filtering~\cite{spantini_coupling_2022, taghvaei_survey_2023} in which nonlinear generalizations to the ensemble Kalman filter (EnKF) have been investigated, we develop a transport-based framework for the purpose of distributed filtering.


Transport methods allow one to sample from a source distribution, such as the posterior in the context of Bayesian inference,  
through the use of transport maps. 
In particular, measure transport characterizes a source distribution as the transformation of a prescribed ``reference'' distribution that is easy to sample (for example, the standard normal). In this work, we 
define the target and reference densities as $\pi$ and $\eta$, respectively. Our aim to seek an invertible transport map $\bm S$ that pushes forward $\pi$ to $\eta$, which we denote by $\bm S_\# \pi = \eta$. The source can then be expressed in terms of the, so-called, pullback distribution $\bm S^\# \eta \coloneqq  (\bm S^{-1})_\# \eta$ using the change of variables formula \cite{villani_optimal_2009,santambrogio_optimal_2015}
\begin{align}
\bm S^\# \eta(\bm z) 
= \eta(\bm S(\bm z)) \det \nabla \bm S(\bm z).
\label{eq:change_of_variable}
\end{align}

Normalizing flows take advantage of the change of variables from a source distribution to a Gaussian reference distribution to enable sampling from a complicated source~\cite{kobyzev_normalizing_2021, papamakarios2021normalizing}. Our goal is similarly motivated to decompose a complicated joint distribution into marginal conditionals, for which we invoke the Knothe Rosenblatt Rearrangement as the map.


\subsection{Conditional Sampling using Triangular Maps} \label{sec:conditioning}

One map that pushes forward a density $\pi$ on $\mathbb{R}^n$ to a standard normal $\eta = \mathcal{N}(0,I_n)$ with useful properties is  
a \textit{monotone} and \textit{triangular} map known as the Knothe-Rosenblatt (KR) rearrangement~\cite{knothe1957contributions, rosenblatt_remarks_1952}. A triangular map $\bm{S}\colon \mathbb{R}^n \to \mathbb{R}^n$ is a multivariate function whose $k$th component only depends only on the first $k$ input variables. That is, $\bm{S}$ can be written as
\begin{align}
\bm S(\bm z) = 
\begin{bmatrix*}[l]
S^1(z_1)\\
\vdots\\
S^n(z_1,z_2,...,z_n)
\end{bmatrix*}.
\label{eq:KRR}
\end{align}
Moreover, we say that the triangular map is monotone if 
\begin{equation}
\xi \mapsto S^k(z_1,\dots,z_{k-1},\xi)  
\end{equation}
is an increasing function for all $z_{1:k-1} \coloneqq (z_1,\dots,z_{k-1}) \in \mathbb{R}^{k-1}$ and $1\leq k\leq n$.

A core property of the KR rearrangement is that its components characterize marginal conditionals of the source and reference distribution. 
First, let us note that given a variable ordering, 
the source density can be factored into a product of marginal conditionals as $\pi = \Pi_k \pi_{Z_k|Z_{1:k-1}}$. Then, for a tensor product reference distribution such as a standard Gaussian, we 
we have that the marginal conditional $\pi_{Z_k|Z_{1:k-1}}(\cdot|z_{1:k-1})$ is pushed forward to the corresponding one-dimensional component of the reference distribution (i.e., the one dimensional standard normal) for all $z_{1:k-1} \in \mathbb{R}^{k-1}$~\cite{santambrogio_optimal_2015}. That is, 
\begin{align}
S^k_\# \pi_{Z_k|Z_{1:k-1}}(\cdot|z_{1:k-1}) = \mathcal{N}(0,1)    
\end{align}

In the context of Bayesian inference, one can use the framework of triangular monotone maps to sample from a conditional distribution $\pi_{\bm X|\bm Y}$ by constructing the map
\begin{equation}
\bm S(\bm y,\bm x) = 
\begin{bmatrix*}[l]
S^1(y_1)\\
\vdots\\
S^d(y_1,...,y_d)\\
S^{d+1}(y_1,...,y_d,x_1)\\
\vdots\\
S^{d+n}(y_1,...,y_n,x_1,...,x_n),\\
\end{bmatrix*}
\end{equation}
which can be partitioned into the following block structure
\begin{align}
\bm S(\bm y,\bm x) = 
\begin{bmatrix*}[l]
\bm S^\mathcal{Y}(\bm y)\\
\bm S^\mathcal{X}(\bm y, \bm x)
\end{bmatrix*},
\end{align}
with maps $\bm S^\mathcal{Y}\colon \mathbb{R}^d \to \mathbb{R}^d$ and $\bm S^\mathcal{X}(\bm y^*, \cdot)\colon \mathbb{R}^n \to \mathbb{R}^n$ for all $\bm y^* \in \mathbb{R}^d$. 

Due to the implicit characterization of conditionals of the joint distribution $\pi_{X,Y}$ by the KR rearrangement and the independence of the components of a standard normal random vector, the map $\bm S^\mathcal{X}$ pushes forward the conditional density $\pi_{\bm X|\bm y^*}(\bm y^*|\cdot)$ to the reference density, i.e.,
\begin{align}
\bm{S}^\mathcal{X}(\bm y^*,\cdot)_\#  \pi_{\bm X|\bm y^*} = \mathcal{N}(0,I_n)
\end{align}
for all $\bm y^* \in \mathbb{R}^{d}$. Equivalently, we note that $\bm S^\mathcal{X}(\bm y^*,\cdot)^{-1}$ pushes forward $\mathcal{N}(0,I_n)$ to $\pi_{\bm X|\bm y^*}$.

To sample from the conditional density $\pi_{\bm X|\bm y^*}(\bm y^*|\cdot)$, we leverage the invertibility of the triangular monotone transport map to map reference samples $\bm{z}^i \sim \eta$ to conditional samples as $\bm{S}^{\mathcal{X}}\left.(\bm{y}^*,\cdot)^{-1}\right|_{\bm{z}^i}$. While the map $\bm{S}^{\mathcal{X}}$ is sufficient for sampling when it is estimated correctly, it can yield inaccurate results otherwise. An alternative transformation $\bm T$ for conditional sampling, that was proposed in~\cite{spantini_coupling_2022}, 
pushes forward $\pi_{\bm Y, \bm X}$ to $\pi_{\bm X|\bm y^*}$ 
using the composed map
\begin{align} \label{eq:prior-to-posterior}
\bm T(\bm y,\bm x) \coloneqq \bm S^\mathcal{X}(\bm y^*,\cdot)^{-1} \circ \bm S^\mathcal{X}(\bm y,\bm x).
\end{align}

\subsection{Estimating Triangular Maps}

To implement the map in~\eqref{eq:prior-to-posterior}, we first build an 
estimator for $\hat{\bm S}^\mathcal{X}$ from a collection of samples $(\bm y^i,\bm x^i) \sim \pi_{\bm Y,\bm X}$. \cite{marzouk_introduction_2016} showed that the components of the map can be computed in parallel, and using convex optimization. Then, we 
define the estimated composed map $\hat{\bm T}(\bm{y},\bm{x}) \coloneqq \hat{\bm S}^\mathcal{X}(\bm y^*,\cdot)^{-1}\circ \hat{\bm S}^\mathcal{X}(\bm{y},\bm{x})$, as is done in \cite{spantini_coupling_2022, ramgraber_ensemble_2022}. 

Let $\mathcal{H}$ be an approximation space of monotone triangular map $U\colon \mathbb{R}^n \to \mathbb{R}^n$ that can be described by finitely many parameters.
The goal of parametric density estimation is to  find a density in the family $(U^\# \eta)_{U \in \mathcal{H}}$ that fits our samples $\bm{z}^1,\dots,\bm{z}^M$. One way to ``fit'' the map to our samples is to minimize the Kullback-Liebler (KL) divergence (relative entropy) between the source distribution $\pi$ and the pullback distribution $\bm{U}^\# \eta$ over the family $\mathcal{H}$, i.e., to minimize $\mathcal{D}_\text{KL}(\pi || \hat{\bm S}^\# \eta)$ \cite{marzouk_introduction_2016, parno_transport_2018}. 
From the relation of KL to maximum likelihood estimation, minimizing $\mathcal{D}_\text{KL}(\pi || \hat{\bm S}^\# \eta)$ is tantamount to finding 
\begin{align}
\hat{\bm S} \in \argmax_{\bm U \in \mathcal{H}} \frac{1}{M}\sum_{i =1}^M  \log  \bm U^\# \eta(\bm z^i).
\label{eq:minimization}
\end{align}

By using equations~\eqref{eq:change_of_variable} and \eqref{eq:KRR}, and choosing our reference density to be the standard normal, we can find each component of $\hat{\bm S}$ in~\eqref{eq:KRR} independently as the minimization of the objective function
\begin{align}
\mathcal{L}^k(U) = \frac{1}{M}\sum_{i =1}^M \left( \frac{1}{2} U^k(\bm z^i)^2 - \log \partial_k U^k(\bm z^i)\right).
\label{eq:objective_function}
\end{align}
That is, $\hat{S}^k \in \argmin_{U^k \in \mathcal{H}_k} \mathcal{L}^k(U)$, where we use the notation $\partial_k U$ to mean the partial derivative of $U$ with respect to its $k$th input variable. Now that we have a simple objective function, the next section explores a simple choice of $\mathcal{H}$. 

\subsection{Affine Ansatz}

In this work we define a family of affine monotone map components of the form
\begin{align}
U^k(\bm z) = u_0 + \sum_{l \leq k} u^k_l z_l, \quad  1 \leq k \leq d + n,
\label{eq:affine_ansatz}
\end{align}
and seek to fit the unknown coefficients $u^k_l$ in accordance with the objective function in~\eqref{eq:objective_function}. Clearly, this yields a triangular map, and monotonicity can be imposed by setting $u^k_l\geq 0$ for $l = k$. We note $\partial_k U^k (\bm z^i)= \partial_{z^i_k} U^k (\bm z^i) = u_k$ and $\partial_{u_l} U^k (\bm z^i)^2 = 2 U^k (\bm z^i)\cdot u_l^k$ for $l \leq k$. So convex optimization is straightforward with the closed-form solution of the gradient of the objective given as
\begin{align}
\frac{d \mathcal{L}^k(U^k)}{d u_l} = \frac{1}{M}\sum_{i =1}^M \ U^k (\bm z^i)\cdot u_l - \frac{\delta_{l,k}}{u_k}.
\end{align}

\section{Data Fusion and Dimension Reduction}
\label{sec:data_fusion}
With the framework of the KR rearrangement, we now show its capability for fusing multiple modes of observations due to the built-in representation of conditional distributions. 

Suppose we have $N$ agents that make $N$ observations $\bm y = (\bm y^1,...,\bm y^N)$, with $\bm y^l \in \mathbb{R}^{d_l}$, of some state $\bm x \in \mathbb{R}^n$. The network topology of the agents is described by a given adjacency matrix $A$, and each agent obtains its neighbors' observations. We concatenate this sequence of observations into vector $\yb{l} \in \mathbb{R}^{\bar{d_l}}$.  Ideally, we would like to sample from the distribution of $\bm{x}$ conditioned on all on observations in the network. However, each agent $l$ is only connected to a subset of the network Nbs($l$).  Now in order to leverage the observations and estimations of each agent in a local manner, we first consider the following triangular map:
\begin{equation}
\label{eq:multi_obs map}
    \bm S(\yb{l}, \bm x) = 
    \begin{bmatrix*}[l]
        \bm S^{\mathcal{Y}}(\yb{l})\\
        \bm S^{\mathcal{X}}(\yb{l}, \bm x)\\
    \end{bmatrix*}.
\end{equation}
This map can be used to define a transformation that takes samples from the joint distribution generated locally at each agent $\pi_{\yb{l},\bm X}$ to the conditional $\pi_{\bm X|\yb{*l}} $ via the composed map $\bm S^{\mathcal{X}}(\yb{*1},..,\yb{*N}, \cdot)^{-1} \circ \bm S^{\mathcal{X}}(\yb{1},...,\yb{N}, \bm x)$. 


To leverage the estimates of neighboring agents, one can employ a consensus algorithm e.g., averaging, gossip, etc. \cite{olfati-saber_consensus_2007}. In this work, we choose a simple averaging algorithm, by which neighboring particle averages $ \hat{\bm x}^{l'} = \frac{1}{M}\sum_{i=1}^M \bm x^{l'}, l' \in \text{Nbs}(l)$ are used to update local particle ensembles $\bm x_1^l,...,\bm x_M^l$ with 
 \begin{align}
     \bm x_i^l \gets \bm x_i^l + \gamma \sum_{l' \in \text{Nbs}(l)} \left( \hat{\bm x}^l - \bm x_i^l\right), \quad i=1,...,M
 \end{align}
and $0<\gamma<1$ is a tunable consensus parameter. It is shown that in a connected network, this update applied to all agents leads to a unique equilibrium when iterated~\cite{olfati-saber_consensus_2007}. In the context of nonlinear filtering, we interleave this consensus pressure on particles with a forecasting step (using the state-space dynamics), and an inference step (using a triangular transport map), as demonstrated in other distributed particle filtering frameworks~\cite{hlinka_distributed_2013}. These 3 stages are presented in Algorithm~\ref{alg:consensus}. The subroutine \textbf{Map} assimilates our predictions with the observed data using an approximate triangular transport map $\hat{\bm T}$. The next section provides one such map $\hat{\bm T}$ that is suited for our proposed filtering scenario.

\subsection{Principal Component Analysis}
It has been shown that in high dimensional state spaces, it may be necessary to regularize the estimation of $\hat{\bm S}$. One approach carried out in~\cite{spantini_inference_2018} imposes sparsity constraint on the functional dependence on 
$\hat{\bm S}$, which exploits weak 
conditional dependence between random variables in each component. 
Another regularization technique is described in Remark 3 of~\cite{spantini_coupling_2022}, in that when assimilating a low-dimensional observation, the particle update should only take place among state components 
that are close to the observed components with respect to a distance function on the indices of the components. 

The approach we choose to take here is similarly motivated, from the observation that the sample covariance matrices $\hat{\Sigma}_{\bm X}, \hat{\Sigma}_{\bm Y}$ for the states and observations, respectively, are low-rank in the numerical experiments below. This is 
especially present with redundant observations in a multi-agent system. Low-rank structure indicates that it may be possible to reduce the complexity of the inference problem by assimilating a reduced number of observations and updating only a low-dimensional subspace of state; see similar analysis and applications of this structure in~\cite{baptista2022gradient, le2022low}. To identify latent spaces for the states and observations with reduced dimensions $q_x < n$ and $q_y < \bar{d_l}$, respectively, we employ principal components analysis (PCA) for these variables separately.

Applying PCA with a set of samples $\bm x_1,..., \bm x_M$ generates an orthonormal matrix $\bm W \in \mathbb{R}^{n \times n}$ such that $\bm W$ maps states $\bm x \in \mathbb{R}^n$ 
as $\bm W^\top \bm{x} \in \mathbb{R}^n$ to a latent space 
that is aligned with the maximum variance directions of the samples. To perform dimension reduction, we map to a lower dimensional latent space using $\bm \hat{\bm W}^\top \bm{x}$ where $\hat{\bm W} \in \mathbb{R}^{n \times q_x}$ contains the first $q_x < n$ columns of $\bm W$. 

Using a similar process, we reduce the dimensions of the observations with the weight matrix $\hat{\bm V} \in \mathbb{R}^{d \times q_y}$, where $\hat{\bm V}$ contains the first $q_y$ columns of the orthonormal matrix given by applying PCA to the $M$ observation samples $\ybi{1}{l},..., \ybi{M}{l}$ (that are generated by sampling from the observation model using $M$ particles $\bm x^l$). 

After identifying the new spaces for the states and observations, we approximate the monotone block-triangular map $\hat{\bm S}$ with the structure
\begin{align}
\hat{\bm S}(\yb{l},\bm x) = 
\begin{bmatrix*}[l]
\hat{\bm S}^{\mathcal{Y}}(\hat{\bm V}^\top \yb{l} )\\
\hat{\bm S}^{\mathcal{X}}(\hat{\bm V}^\top \yb{l} , \hat{\bm W}^\top \bm{x}),\\
\end{bmatrix*}
\end{align}
with maps $\hat{\bm S}^{\mathcal{Y}}\colon \mathbb{R}^{q_y} \to \mathbb{R}^{q_y}$ and $\hat{\bm S}^{\mathcal{X}}(\hat{\bm V}^\top\yb{*l}, \cdot)\colon \mathbb{R}^{q_x} \to \mathbb{R}^{q_x}$ for all $\yb{*l} \in \mathbb{R}^d$. Following the approach in Section~\ref{sec:conditioning}, we define the estimated composed map as
\begin{align*}
\hat{\bm T}(\bm x, \yb{l}) \coloneqq 
\hat{\bm W} \left(\bm \hat{\bm S}^{\mathcal{X}}(\hat{\bm V}^\top \yb{*l}, \cdot)^{-1} \circ \hat{\bm S}^{\mathcal{X}}(\hat{\bm V}^\top \yb{l}, \hat{\bm W}^\top \bm{x}) \right).
\end{align*}

We note that the estimated weight matrices $\hat{\bm W}$ and $\hat{\bm V}$ are updated at each iteration of the filtering process, thereby enabling us to update all components of the state process by focusing the inference 
on subspaces of the states and observations with the greatest variance. 
The process of assimilating one agent's estimates $\bm x_1^l, ..., \bm x_M^l$ with its neighboring observations $\{ \bm h^{l'}\}_{l' \in \text{Nbs}(l)}$ using PCA is given by Algorithm~\ref{alg:PCAMap} (which can be used as the \textbf{Map} subroutine in  Algorithm~\ref{alg:consensus}). We note that the subroutine \textbf{PCA} simply takes a collection of samples and returns the truncated weight matrix $\hat{\bm W}$ with $q$ columns that is associated with the data's principal components.

\begin{algorithm}
\caption{(\textbf{ConsensusFiltering}) Given $N$ agents, connected by network with adjacency matrix $A$, state-space model \ref{eq:dynamics}, \ref{eq:observation}, particles $\{\bm x_i^l \}_{i=1}^M$ are updated through a forecast step, an assimilation step, and a consensus step.}
\label{alg:consensus}
\begin{algorithmic}
\For{$l \gets 1:N$}
\State $\bm x_i^l \gets $ \textbf{Forecast} ($\bm x_i^l$)
\EndFor
\For{$l \gets 1:N$}
\State $\{\bm x_i^l \}_{i=1}^M \gets$ \textbf{Map} $\left(\{\bm x_i^l \}_{i=1}^M, \{\bm y^{*l'}, \bm h^{l'}\}_{l' \in \text{Nbs}(l)} \right)$
\EndFor
\For{$l \gets 1:N$}
\State $\{\bm x_i^l \}_{i=1}^M \gets $ \textbf{Consensus} $\left([\{\bm x_i^{l'} \}_{i=1}^M]_{l' \in \text{Nbs}(l)} \right)$
\EndFor 
\end{algorithmic}
\end{algorithm}

\begin{algorithm}
\caption{(\textbf{PCAMap}) Given $q_x \leq \bar{d_l}$, $q_y \leq N$, $M$ samples $\bm x_1^l,...., \bm x_M^l$ belonging to agent $l$, neighboring observation functions $\bm h^{(l')}$ and observations $\yb{*l}$, we update particles $\bm x_1^l,...., \bm x_M^l$ according to a transport map over principal components in order to sample from the posterior distribution $\pi_{\bm x |\bm y^{*l}}$}
\label{alg:PCAMap}
\begin{algorithmic}[1]
\For{$i \gets 1:M \do$} 
\State $\ybi{i}{l} \gets  \{\bm h^{(l')}(\bm x_i^l, \bm \xi)\}_{l' \in \text{Nbs}(l)}$ Sample observations
\EndFor
\State $\hat{\bm V} \gets $ \textbf{PCA}($\{\ybi{i}{l}\}_{i=1}^M, q_y$) 
\State $\hat{\bm W} \gets $ \textbf{PCA}($\{\bm x^l_i\}_{i=1}^M, q_x$) 
\State $\hat{\bm S}^{\mathcal{X}} \gets $ Estimate the KR-rearrangement in \ref{eq:minimization} using samples $(\hat{\bm V}^\top \bm y^l_i, \hat{\bm W}^\top \bm x_i^l)_{i=1}^M$
\For{$i \gets 1:M \do$} 
\State $\bm x_i \gets \hat{\bm W} \left(\hat{\bm S}^{\mathcal{X}}(\hat{\bm V}^\top \bm y^{*l}, \cdot)^{-1} \circ \bm \hat{\bm S}^{\mathcal{X}}(\hat{\bm V}^\top \bm y^l_i, \hat{\bm W}^\top \bm x_i^l) \right)$
\EndFor
\State \Return $\bm x_1^l,..., \bm x_M^l$
\end{algorithmic}
\end{algorithm}

\section{Case Study}
\label{sec:case_study}
\subsection{Clohessy Wiltshire Hill Equations}
When studying spacecraft in close proximity, it is common to leverage the relative Hill's frame \cite{hill_researches_1878} to characterize the motion of satellites in network~\cite{blondin_robust_2022}. In particular, the relative position of a deputy relative to the chief is described in the Hill frame by the three unit vectors: $e_R$ is pointing radially outward from the the center of the orbit, $e_T$ is along the orbit track of the target body, and $e_N$ is along the orbital angular momentum vector of the target body. In the case of a chief satellite in a circular orbit with deputy satellites moving about the chief in an elliptical orbit, the translational relative dynamics are given in the Hill frame by the Clohessy-Wiltshire equations \cite{clohessy_terminal_1960} with disturbances modelled by additive white noise $\eta(t)$ with standard deviation $\sigma$. That is, 
\begin{align}
\dot{\bm x}_T =
\begin{bmatrix}
 &0 &0 &0 &1 &0 &0\\
 &0 &0 &0 &0 &1 &0\\
 &0 &0 &0 &0 &0 &1\\
&3\alpha^2 &0 &0 &0 &2\alpha &0\\
 &0 &0 &0 &-2\alpha &0 &0\\
 &0 &0 &-\alpha^2 &0 &0 &0\\
\end{bmatrix} \bm x_T + \sigma \eta(t),
\label{eq:clohessy_lti}
\end{align}
where $\alpha = \sqrt{\mu/a^3}$ is the orbital rate of the chief body and the state vector 
is decomposed as $\bm x_T = (x,y,z,\dot{x},\dot{y},\dot{z})$.

One can similarly obtain nonlinear relative attitude dynamics for the deputy in the Hill frame. First we define $[\cdot]_\times:\mathbb{R}^3 \to \mathfrak{so}(3)$ as the cross product operator
\begin{align}
[\bm \omega]_\times =
\begin{bmatrix}
    0 &-\omega_3 &\omega_2\\
    \omega_3 &0 &-\omega_1\\
    -\omega_2 &\omega_1 &0
\end{bmatrix}.
\end{align}
The dynamics with additive white noise are
\begin{align}
    \dot{\bm \omega}^\mathcal{D} &= \dot{\bm \omega}^\mathcal{E} + [\bm \omega^\mathcal{D}]_\times \mathcal{R} \bm \omega^\mathcal{O} + \sigma \eta(t),
\end{align}
where $\bm \omega^\mathcal{D}$ is the angular velocity of the deputy frame relative to the Hill frame, $\bm \omega^\mathcal{E}$ the angular velocity of the deputy frame relative to Earth inertial frame, $\bm \omega^\mathcal{O}$ is the angular velocity of the Hill frame relative to Earth inertial frame, and $\mathcal{R}$ is the rotation matrix that transforms the Hill frame to the deputy frame. To recover the orientation of the deputy in the Hill frame parameterized as a unit quaternion $\bm q = (q_1, q_2, q_3, q_4)$, we employ the equation
\begin{align}
    \begin{bmatrix}
        \dot{q}_1\\
        \dot{q}_2\\
        \dot{q}_3\\
        \dot{q}_4
    \end{bmatrix} &= - \frac{1}{2}
    \begin{bmatrix}
        -q_2 &-q_3 &-q_4\\
        q_1 &q_4 &-q_3\\
        -q_4 &q_1 &q_2\\
        q_3 &-q_2 &q_1
    \end{bmatrix}
    \begin{bmatrix}
        \omega^\mathcal{D}_1\\
        \omega^\mathcal{D}_2\\
        \omega^\mathcal{D}_3\\
    \end{bmatrix}.
\label{eq:clohessy_quat}
\end{align}
We ultimately aim to estimate the combined translation $\bm x_T$ and attitude $\bm x_R := (q_1, q_2, q_3, q_4, \omega^\mathcal{D}_1, \omega^\mathcal{D}_2, \omega^\mathcal{D}_3)$, with $\bm x = (\bm x_T, \bm x_R)$.

\subsection{Numerical Results}
Two dynamical models salient to multi-agent space operations: the CW translation equations~\eqref{eq:clohessy_lti} (linear) and the full CW equations~\eqref{eq:clohessy_quat} (nonlinear), are considered for testing our proposed scheme. 
The simulation of the ground truth states is conducted by numerically integrating the ODE using SciPy's implementation of the fourth-order Runge-Kutta method~\cite{2020SciPy-NMeth} with time step $10^{-2}$. Observations of the state are made with time step $10^{-1}$ from time $t=0$ to $t=20$ with four different observation functions considered in Table~\ref{table:observations}. Process noise is Gaussian with independent marginals of mean zero and standard deviation $\sigma = 0.2$. The orbital rate is set to $\alpha = 0.1$, with orbit $8.413\times 10^6m$ and orbit period $7680s$. Initial conditions are set to $\bm x_0 = (x=100,y=0,z=0,\dot{x}=10,\dot{y}=0.1,\dot{z}=0,\omega_x=0.3,\omega_y=0.2,\omega_z=0.1,q_1=1,q_2=0,q_3=0,q_4=0)$. We choose $M=200$ particles with initial particle distribution independently and identically distributed $\bm X_T \sim \mathcal{N}(\bm x_0 + 3, 25\cdot \bm I_6)$ and $\bm X_R \sim \mathcal{N}(\bm x_0 + 0.1, 0.04\cdot \bm I_7)$. 

Triangular maps are parameterized as in~\eqref{eq:affine_ansatz} and they are approximated using the ECOS convex optimization solver implementation in CVXPY \cite{agrawal2018rewriting}. The optimization is run until a maximum number of $10^4$ iterations or the residual reaches a threshold of $10^-5$. The code to reproduce the results is available online~\footnote{\url{https://github.com/Dan-Grange/DistributedCoupling}}.
 
\begin{figure}[t]
    \centering
    \includegraphics[width=0.45\textwidth,clip]{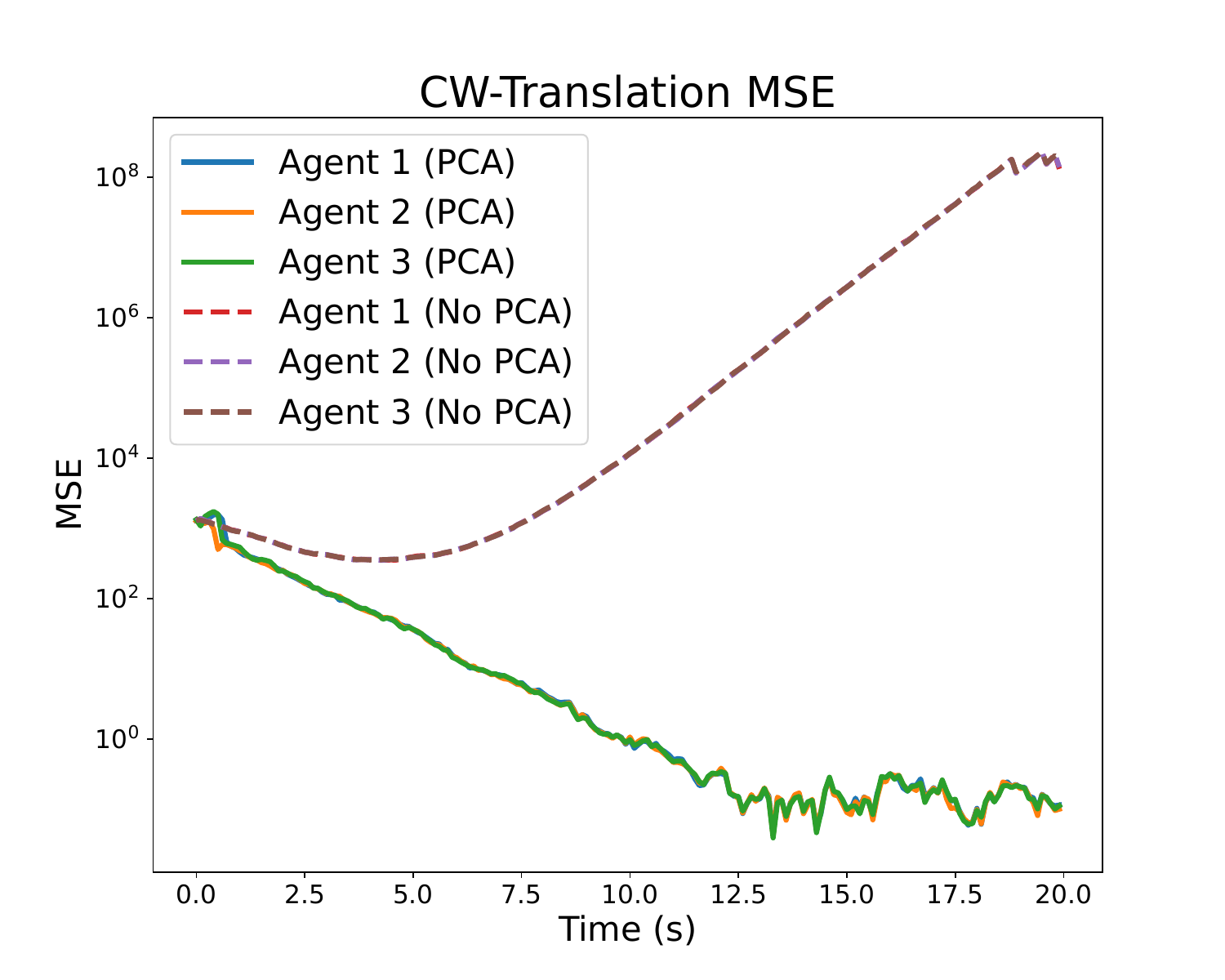}
    \caption{Simulation results of Algorithm \ref{alg:consensus} where the mapping step is performed either using PCA (solid lines) or without PCA (dashed lines). Agents are setup according to Table~\ref{table:1}. The ensemble average $\hat{\bm x}^l = \frac{1}{M}\sum_{i=1}^M \bm x_i^l$ at each time step for each agent is measured against the true value $\bm x_i$ using the mean squared error (MSE) $\|\hat{\bm x} - \bm x\|_2^2$. We observe that without PCA, there is unstable filter performance with $M=200$.}
    \label{fig:translation}
\end{figure}

In the first assessment of our proposed filter given by Algorithm~\ref{alg:consensus}, we consider the linear model of the CW translation equation~\eqref{eq:clohessy_lti}. Three agents make complimentary observations of the 6-dimensional state $\bm x_T$. PCA parameters $q_y=3$ and $q_x=4$ are chosen for dimension reduction, and additional parameters for the agent network are given in Table \ref{table:1}. Figure~\ref{fig:translation} shows the average mean squared error (MSE) when using either PCA dimension reduction, or not using dimension reduction. Without PCA dimension reduction, we notice unstable filtering performance with low-sample sizes.

\begin{figure}[t]
    \centering
    \includegraphics[width=0.45\textwidth,clip]{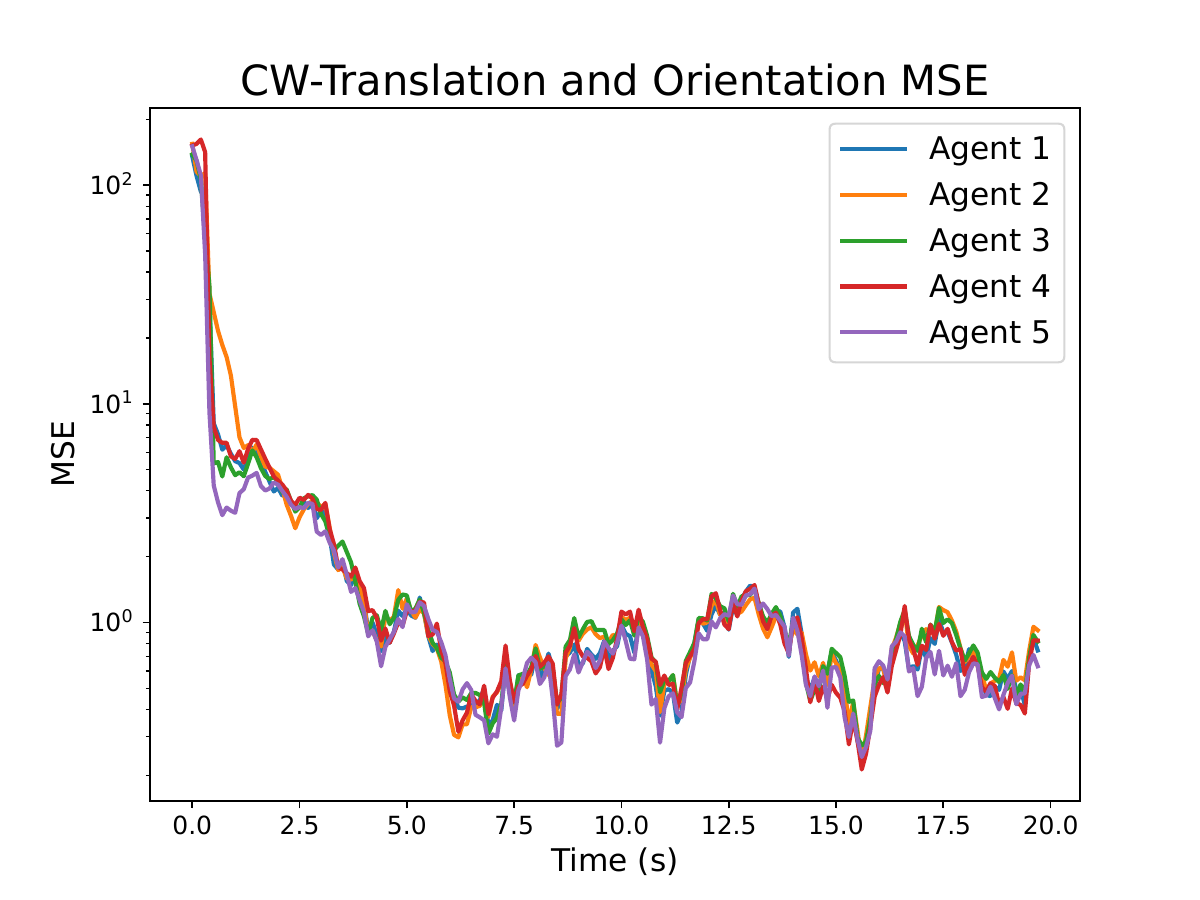}
    \caption{MSE results for Algorithm \ref{alg:consensus} using PCA mapping with agent setup given in Table~\ref{table:2}.} 
    \label{fig:full CW}
\end{figure}

For our second assessment, we consider the nonlinear model of the full CW equations. Five agents make complimentary observations of the 13-dimensional state $\bm x = (\bm x_T, \bm x_R)$, where $\bm x_T$ evolves according to~\eqref{eq:clohessy_lti} and $\bm x_R$ evolves according to~\eqref{eq:clohessy_quat}. PCA dimension reduction parameters are set to $q_y=6$ and $q_x=10$, and parameters for the network are given in Table \ref{table:2}. Figure~\ref{fig:full CW} shows the average mean squared error (MSE) for the scheme, with similar agreement between the agents, which is expected under consensus. 

\begin{table}[!ht]
\centering
\begin{tabular}{lllr}
\toprule
Obs. Type & Obs. Function \\
\midrule
direct&       $h(\bm x) = \bm x$     \\
differential& $h(\bm x^{(i)}, \bm x^{(i-1)}) = \bm x^{(i)} - \bm x^{(i-1)}$  \\
rangefinder&  $h(\bm x) = \|\bm x \|_2$ \\
angle&  $h(x,y) = \arctan(y/x) $ \\
\bottomrule
\end{tabular}
\caption{Linear and nonlinear observation functions correspond ingto sensors such as a laser range finder, camera, event camera, or LIDAR.}
\label{table:observations}
\end{table}

\begin{table}[!ht]
\centering
\begin{tabular}{lllrr}
\toprule
Agent & Obs. Dimensions & Obs. Type &  Noise & Neighbors \\
\midrule
A&              [0, 1] &           direct &    0.2 & [A,B,C]\\
B&              [1, 2] &           angle &     0.2 & [A,B]\\
C&           [3, 4, 5] &           direct &    0.1 & [A,C]\\
\bottomrule
\end{tabular}
\caption{Agent parameters for linear translation state model.}
\label{table:1}
\end{table}

\begin{table}[!ht]
\centering
\begin{tabular}{lllrr}
\toprule
Agent & Obs. Dimensions & Obs. Type &  Noise & Neighbors\\
\midrule
A &           [0, 1, 2] &           direct &    0.2& [A,B,C,D,E] \\
B &           [3, 4, 5] &     differential &    0.1& [A,B] \\
C &           [0, 1, 2] &            range &    0.1& [A,C] \\
D &           [6, 7, 8] &           direct &    0.1& [A,D] \\
E &     [9, 10, 11, 12] &           direct &    0.1& [A,E] \\
\bottomrule
\end{tabular}
\caption{Agent parameters for nonlinear translation and orientation state model.}
\label{table:2}
\end{table}

\section{Conclusion and Future Work}
We have demonstrated through a case study for the simulation of space vehicles that distributed state estimation of the relative position and relative attitude can be achieved through via triangular transport maps. Current efforts to improve results involve extending more ``expressive'' transport maps to the multi-agent setting. Another thrust is in the construction of a global transport map that shares information across agents. In the absence of communication barriers between agents, all particles can serve as samples from which a global transport map can be found by solving an optimization problem, i.e.,
$
    \min_U \sum_{l=1}^N \mathcal{L}_l(U).
$
Moreover, we can separate the optimization across agents with the introduction of a consensus constraint over $U$. That is, we solve 
\begin{align}
    &\text{minimize } \sum_{l=1}^N \mathcal{L}(U_l)\\
    &\text{subject to }  U_l = Z, \ l=1,..,N, \nonumber
    \label{eq:global_consensus_problem}
\end{align}
where $Z$ is the putative 'global' model. Future work will investigate methods to approach this optimization problem with the specialized alternating direction method of multipliers (ADMM) \cite{boyd_distributed_2010}. 

\bibliographystyle{IEEEtran.bst}
\bibliography{IEEEabrv.bib,ACC_refs.bib}

\end{document}